\documentclass[12pt,preprint]{aastex}

\shorttitle{The Nature of Part Time Pulsars} \shortauthors{Li}

\begin{document}


\title{On the Nature of Part Time Radio Pulsars}


\author{Xiang-Dong Li}
\affil{Department of Astronomy, Nanjing University,
    Nanjing 210093, China;\\ lixd@nju.edu.cn}



\begin{abstract}

The recent discovery of rotating radio transients and the
quasi-periodicity of pulsar activity in the radio pulsar PSR
B1931$+$24 has challenged the conventional theory of radio pulsar
emission. Here we suggest that these phenomena could be due to the
interaction between the neutron star magnetosphere and the
surrounding debris disk. The pattern of pulsar emission depends on
whether the disk can penetrate the light cylinder and efficiently
quench the processes of particle production and acceleration
inside the magnetospheric gap. A precessing disk may naturally
account for the switch-on/off behavior in PSR B1931$+$24.

\end{abstract}


\section{Introduction}

A small group of rotating radio transients (RRATs) were recently
reported by \citet{mcl06}. These objects are characterized by
single, dispersed bursts of radio emission with durations between
2 and 30 ms. The average time intervals between bursts range from
4 minutes to 3 hours, with radio emission typically detectable for
$< 1$ s per day. Periodicities in the range $0.4-7$ s for 10 of
the 11 sources have been measured, suggesting that they are
rotating neutron stars. Three of the sources have measured period
derivatives, with one (RRAT J1819$-$1458) having a very high
inferred magnetic field of $5\times 10^{13}$ G, if spin-down by
magnetic dipole radiation is assumed.

A similar bursting radio source GCRT J1745$-$3009 was detected
previously \citep{hym05}, whose notable properties include
``flares" approximately 1 Jy in magnitude lasting approximately 10
minutes each and occurring at apparently regular 77 minute
intervals. GCRT J1745$-$3009 is located approximately $10\arcmin$
from the Galactic center, and just outside of the shell-type
supernova remnant (SNR) G359.1$-$0.5 \citep{rei84}. If GCRT
J1745$-$3009 and the SNR are related, then GCRT J1745$-$3009 would
have the age of $\sim 10^5$ yr, comparable to RRAT J1819$-$1458.
But the overall age distribution of RRATs is still not clear,
although 5 out of 10 have spin periods $P>4$ s.

Despite the small number of RRATs detected, \citet{mcl06} suggest
that their ephemeral nature may point to a total Galactic
population significantly exceeding that of the regularly pulsing
radio pulsars. Investigation of the nature of RRATs will be of
great interest on understanding pulsar formation, evolution and
radiation mechanisms.

Several possible interpretations for the peculiar properties of
RRATs as well as GCRT J1745$-$3009 have been examined.  For the
latter, models involving a precessing pulsar \citep{zhu06}, binary
neutron star \citep{tur05}, and transient white dwarf pulsar
\citep{zha05} have been suggested. More recently, \citet{zha06}
discuss two possible interpretations to RRATs: the first model
suggests that these objects are pulsars slightly below the radio
emission ``death line", and become active occasionally when the
conditions for pair production and coherent emission are
satisfied; the second one invokes a radio emission direction
reversal in normal pulsars due to some unknown reasons.

The re-activated dead pulsar model in Zhang et al. (2006) is
particularly interesting, although the so-called ``death line" is
highly uncertain, because it depends not only on the magnetic
field configuration, but significantly on the origin of gamma
quanta, which are responsible for pair production. The outbursts
of radio emission are assumed to be caused by internal magnetic
field evolution in the neutron stars - when stronger multipole
magnetic fields emerge to the polar cap region of the neutron
star, the pair production condition could be satisfied and the
neutron star behaves like a radio pulsar. A few lines of
implications can be drawn from this idea. First, the large
population of RRATs may be either due to higher birthrate of young
pulsars than previously thought, or caused by a pile-up effect at
large periods. This could be testified by future population
synthesis of radio pulsars. Second, since reconnection typically
occurs at a fraction of Alf\'ven velocity $v_{\rm A}$, in the case
where the instability is driven by the internal field, the growth
time of the instability is \citep{tho95}
\begin{equation}
\Delta t\sim \frac{\Delta l}{v_{\rm A}}\sim
10B_{*,12}^{-1}\rho_{15}^{1/2}(\frac{\Delta l}{0.1R_*})\,{\rm s},
\end{equation}
where $\Delta l$ is the displacement of the field lines,
$B_*=10^{12}B_{*,12}$ G, $\rho=10^{15}\rho_{15}$ gcm$^{-1}$, and
$R_*$ the surface magnetic field strength, the core density, and
the radius of the neutron star, respectively. To be compared to
the measured RRAT burst duration of $\sim 10$ ms, one has to
assume that the field readjustment should take place in the outer
crust of the neutron star, where $\rho_{15}\sim 10^{-6}$.

In this {\it Letter} we present an alternative interpretation to
RRATs. We propose that some of them may be isolated neutron stars
surrounded by a debris disk, which originate either from the
supernovae that produced the neutron stars or from the captured
interstellar medium. The neutron stars act as a propeller when the
disk penetrates inside the light cylinder, and the outflow or wind
from the disk may quench the pair production processes in the
pulsar magnetosphere with only transient radio emission allowed.
The interaction between a processing debris disk with the neutron
star magnetic field may also be responsible for the quasi-periodic
switch-on/off transition in PSR B1931$+$24 recently reported by
\citet{kra06}.

\section{RRATs}

In the standard model for radio pulsars, the rotationally induced
electric field of a rotating, magnetized neutron star pulls the
plasma off the surface and eject it beyond the light cylinder to
form a relativistic wind \citep{gol69}. Pulsar emission is then
associated with the acceleration of particles to maintain this
wind, which appears by the avalanche process of (e$^{\pm}$) pair
production \citep{rud75}. In the polar-cap models, the
acceleration and radiation occur near the magnetic poles in the
inner magnetosphere, while in the outer-gap models, these
processes occur in the outer magnetosphere (see Kaspi, Roberts, \&
Harding for a review). The flow of the plasma along open field
lines results in some plasma void regionsin the vicinity of
null-charge surfaces. In such charge-deficient regions (or
``gaps") electric field parallel the magnetic field lines
$E_{||}\neq 0$ is sustained, electrons and positrons can be
accelerated to relativistic energies.

It is interesting to see whether the conditions for pulsar
emission will be satisfied if there is a disk surrounding the
neutron star. Michel \& Dessler (1981, 1983; see also Michel 1988)
argued that radio pulsars and X-ray pulsars differ mainly in the
fact that the former are surrounded by a supernova fallback disk
with negligible accretion, while the latter are surrounded by an
accretion disk. Alternatively, the debris disks around isolated
neutron stars may result from the captured interstellar medium
\citep[e.g.][]{pop00}. The fallback disk model was recently
adopted to account for the observational characteristics of
anomalous X-ray pulsars \citep{yus95,cha00,alp01}. Such disks can
also influence pulsar braking indices and timing ages.
\citet{men01} have explored a disk model for the spin-down of
young radio pulsars, in which the neutron star loses rotational
energy not only by emitting magnetic dipole radiation, but also by
torquing a supernova fallback disk. \citet{mar01} considered a
similar model to explain the discrepancies between a pulsar's true
age and its characteristic age. Recent X-ray observations also
show that some young pulsars, such as the Crab and Vela pulsars,
may have the jet configuration, which suggests the existence of a
disk surrounding the neutron star \citep[][and references
therein]{bla04}. The strongest constraints on the presence of a
disk is given by its optical and longer wavelength emission.
\citet{perna00} have examined the reprocessing of beamed pulsar
emission by the debris disks. They found that, since the
reradiated flux gives the dominant contribution at long
wavelengths produced in the bulk of the disk, whereas the optical
emission is generated in its innermost part, the optical rather
the longer wavelength emission would be highly suppressed, if the
inner edge of the disk were truncated at a radius larger than the
magnetospheric radius. Most recently, \citet{wan06} report the
discovery of mid-infrared emission from a cool disk around the
isolated young X-ray pulsar 4U 0142$+$61, presenting the first
direct evidence for supernova fallback.

In the original picture of \citet{stu71} and \citet{mic81} the
debris disk is magnetically coupled with the neutron star with no
accretion. This may be true for cold disks with extremely low
viscosity. However, it is conventionally thought that the pulsar
emission will quenched if the disk wind plasma penetrating into
the light cylinder.

Most of the neutron stars with a surrounding disk should have
experienced the accretor and propeller regimes \citep{ill75}. The
existence of a disk inside the light cylinder may significantly
influence the pulsar radiation processes. Magnetocentrifugally
driven outflows from the disks has been discussed by a number of
authors during the propeller phase
\citep[e.g.][]{cam90,kon91,lov95,aga00,ust06}. The disk wind
itself may also be strong enough to influence the structure of
pulsar winds \citep{bla04}. It has already been shown that the
$\sim 10^{12}$ V potential difference across the magnetospheric
gap and the outward-directed electric field required by the
Ruderman-Sutherland model for the generation of radio waves will
be negated, if the number density of matter at the Alf\'ven
surface is greater than $\sim 7.2\times 10^7$ cm$^{-3}$
\citep{wan83}, a condition satisfied by most neutron stars with a
debris disk where there is significant wind or outflow from the
disk. The density of the outflow plasma, if similar to that in the
disk, can be estimated to be
\begin{equation}
\rho_{\rm w}=\frac{\dot{M}}{2\pi RHv_{\rm r}m_{\rm H}}\simeq
2.2\times 10^{15}(\frac{\dot{M}}{10^{14}{\rm
gs}^{-1}})(\frac{\alpha}{0.01})^{-1}(\frac{H/R}{0.1})^{-3}(\frac{R}{10^9\,{\rm
cm}})^{-3/2}\,{\rm cm}^{-3},
\end{equation}
where $\dot{M}$ is the mass inflow rate in the disk, $H$ the half
thickness, $v_{\rm r}$ the radial velocity, $R$ the inner radius
of the disk, $\alpha$ the viscosity parameter, and $m_{\rm H}$ the
proton mass, respectively. It can be much larger than the
Goldreich-Julian density $\rho_{\rm GJ}$ for typical values of the
adopted parameters, if one assumes that the electron-positron
pairs in the gap are close to saturation,
\begin{equation}
\rho_{\rm GJ}=\frac{\Omega B}{2\pi ce}=7\times 10^{10}B_{*,
12}P^{-1}(\frac{R_*}{R})^{-3}\,{\rm cm}^{-3},
\end{equation}
where $P$ is the neutron star spin period. Failure of pulsar
emission may also partly result from the fact that coherent
radiowaves with a wavelength longer than 75 cm can be absorbed
effectively in the wind plasma \citep{ill75}.

The flow in the inner part of the accretion disk is expected to
have density fluctuations (``clumps") produced by a variety of
mechanisms, such as thermal instability, Kelvin-Helmholtz
instability, and magnetoturbulence \citep[see][]{lam85,shi87}. The
clumpy wind density would be much higher than the averaged value
estimated above. They may also leave short, sporadic
``transparent" time for the development of particle acceleration
in the gap and generation of pulsar emission. If we assume that
the typical clump separation is less than the disk height $H_{\rm
in}$ at the inner edge $R_{\rm in}$ of the disk, the duration of
successful pulsar emission should be less than
\begin{equation}
\tau\sim H_{\rm in}/v_{\rm esc},
\end{equation}
where $v_{\rm esc}$ is the escape velocity at $R_{\rm in}$. If
$R_{\rm c}<R_{\rm in}<R_{\rm lc}$ and $H/R\la 0.1$,  we have $11P$
ms $<\tau<1.7P^{3/2}$ s. Here $R_{\rm c}\equiv(GMP^2/4\pi)^2$ is
the corotation radius and $R_{\rm lc}\equiv cP/2\pi$ is the light
cylinder radius, respectively. As pointed out by Zhang et al.
(2006), the dynamical time scale of the inner gap ($\sim h_{\rm
gap}/c\sim 10^{-6}-10^{-4}$ s, where $\sim h_{\rm gap}$ is the
height of the gap) is much smaller than the rotation period $P$.
So the time scale to develop a pair cascade is much shorter than
$\tau$. Its magnitude seems compatible with the burst durations
measured so far.

The debris disk may be popular in relatively young neutron stars.
Jiang \& Li (2005) performed Monte-Carlo simulation of pulsar
evolution, assuming that all neutron stars are born with a
surrounding supernova fallback disk with the initial masses of the
disk ranging from $10^{-6}\,M_{\sun}$ to $10^{-2}\,M_{\sun}$. They
found that the emerging proportion of disk-fed neutron stars (i.e.
with the disk extending inside the light cylinder) is $\sim
20\%-50\%$ at age of $10^3$ years, and $\sim 10\%-25\%$ at age of
$10^4$ years. Obviously these numbers are sensitive to the
assumptions for the initial parameters, e.g., the distributions of
the initial disk masses, of the neutron star spin periods,
magnetic fields, and most importantly, the mechanisms of the
propeller spin-down. However, it clearly demonstrates that a
considerable fraction of isolated neutron stars could harbor a
debris disk with sufficiently long time (\citet{pop00} suggested
that a fraction of $0.1\%-0.2\%$ of all isolated neutron stars may
be presently in the propeller stage due interaction with the
interstellar medium). It was also found that the ratio of the
characteristic age $t_{\rm c}=P/2\dot{P}$ and the true age $t$
distributes within a relatively wide range from $\sim 0.1$ to
$\sim 10$, indicating that $t_{\rm c}$ and the magnetic field
strength estimated from magnetic dipole radiation may considerably
deviate from the actual values for these neutron stars. The
disk-assisted spin-down may also explain why RRATs have relatively
long spin periods compared with normal isolated radio pulsars.

\citet{rey06} recently report the discovery of the X-ray
counterpart to RRAT J1819$-$1458. While their data are
insufficient for fitting to more detailed neutron star atmosphere
models, they suggest that the emission from RRAT J1819$-$1458 is
consistent with a cooling neutron star of age $\sim 10^4-10^5$ yr,
at a distance $\la 2$ kpc. This seems to be in contradiction with
our scenario since thermal, soft radiation is not expected from a
propeller, in which nonthermal magnetospheric emission should
dominate (e.g. Popov, Turolla, \& Possenti 2006). From the work of
\citet{can90} and \citet{min97} for supernova fallback, we can
roughly estimate the mass inflow rate in the disk as
\begin{equation}
\dot{M}\simeq 1.8\times 10^{14}(\frac{M}{1
M_{\odot}})(\frac{\Delta
M}{10^{-5}M_{\odot}})(\frac{\alpha}{0.01}) (\frac{t}{10^5\,{\rm
yr}})^{-1.35}\,{\rm gs}^{-1},
\end{equation}
where $\Delta M$ is the amount of fallback material. With Eq.~(5)
the maximum luminosity released by the propeller process is
\begin{equation}
L_{\rm prop}=\frac{GM\dot{M}}{R_{\rm m}}\simeq 2.4\times
10^{30}(\frac{M}{1 M_{\odot}})^2(\frac{\Delta
M}{10^{-5}M_{\odot}})(\frac{R_{\rm m}}{0.5R_{\rm
lc}})^{-1}(\frac{\alpha}{0.01}) (\frac{t}{10^5\,{\rm
yr}})^{-1.35}\,{\rm ergs}^{-1},
\end{equation}
for RRAT J1819$-$1458, which is much smaller than the measured
luminosity $\sim 10^{33}\,{\rm ergs}^{-1}$, implying that neutron
star cooling could still dominate X-ray emission in this object.
But we mention that a neutron star undergoing propeller spindown
could be a weak point source of $\gamma$-ray radiation during the
(radio-)quiescent state \citep{wan85}.

\section{PSR B1931$+$24}

More recently \citet{kra06} report the quasi-periodical pattern in
the radio pulsar PSR B1931$+$24: the radio emission switches off
in less than 10 seconds after the active phases of $\sim 5-10$
days, and remains undetectable for the next $\sim 25-35$ days when
it switches on again. More remarkably, the pulsar rotation slows
down $50\%$ faster when it is on than when it is off, indicating
an increase in magnetospheric currents when the pulsar switches
on. As pointed out by \citet{kra06}, the discovery of PSR
B1931+24's behaviour suggests that many more such objects exist in
the Galaxy, and the bursting radio source GCRT J1745$-$3009 may
turn out to be a short-timescale version of PSR B1931$+$24 and
hence to be a radio pulsar.

The $35$ day period is not likely to be attributed to free
precession of the neutron star, because no evidence of expected
profile changes is found \citep{kra06}. However, it might be
accounted for in our pulsar $+$ debris disk model. Here we suggest
that the $35$ day period is the precession period  of the debris
disk. It is well known that the neutron star receives a kick
during the supernova explosion, and it is likely that there is
misalignment between the angular momenta of the fallback disk and
the neutron star, leading to free precession of the disk
\citep{kat73,rob74}. The disk precession can also be induced by
the radiation or magnetic torques generated from the neutron star
\citep[e.g.][]{pet77,pri96,lai99}. There is extensive evidence for
a warped, precessing disk in X-ray binaries and active galactic
nuclei \citep[e.g.][and references therein]{ogi01}.

The quasi-periodicity in PSR B1931$+$24 may be explained in the
following picture. We assume that the inner edge of the debris
disk is close to the pulsar's light cylinder. It is known that the
horizontal distance of the disk from the spin axis of the neutron
star always changes during the precession. As soon as the disk
penetrates inside the light cylinder, the propeller process
commences along with outflows from the disk, particle acceleration
processes in the magnetospheric gap are then quenched and the
coherent radio emission cuts off. The neutron star slows down only
by magnetic dipole radiation. The pulsar radiation switches on
when the disk moves outside the light cylinder. In this case both
magnetic dipole radiation and pulsar wind brake the neutron star,
so that the pulsar slows down faster than during the off phase.
This scenario also suggests that PSR B1931$+$24 may appear as a
RRAT during the off phase. The recent detection of transient
pulsed radio emission from the anomalous X-ray pulsar XTE
J1810$-$197 \citep{cam06} could be an example of this transition.

%\section{Conclusions}

\acknowledgments {I am grateful to Prof. M. Ruderman for helpful
discussion. This work was supported by the Natural Science
Foundation of China under grant number 10573010.}


\begin{thebibliography}{}
\bibitem[Agapitou \& Papaloizou(2000)]{aga00} Agapitou, V. \& Papaloizou, J. C. B.
2000, MNRAS, 317, 273
\bibitem[Alpar(2001)]{alp01} Alpar, M. A. 2001, ApJ, 554, 1245
\bibitem[Blackman \& Perna(2004)]{bla04} Blackman, E. G. \& Perna, R. 2004, ApJ,
601, L71
\bibitem[Camenzind(1990)]{cam90} Camenzind, M. 1990, Rev. Mod. Astron., 3,
234
\bibitem[Camilo et al.(2006)]{cam06} Camilo, F. et al. 2006,
astro-ph/06605429
\bibitem[Cannizzo, Lee, \& Goodman(1990)]{can90} Cannizzo, J. K., Lee, H. M., \& Goodman, J., 1990, ApJ, 351, 38
\bibitem[Chattterjee, Hernquist, \& Narayan(2000)]{cha00} Chattterjee, P., Hernquist, L., \& Narayan, R., 2000, ApJ, 534,
373
\bibitem[Goldreich \& Julian(1969)]{gol69} Goldreich, P. \& Julian, W. H. 1969, ApJ, 157, 869
\bibitem[Hyman et al.(2005)]{hym05} Hyman, S. D., Lazio, T. J. W., Kassim, N. E., Ray, P. S., Markwardt, C. B.
\& Yusef-Zadeh, F. 2005, Nat, 434, 50
\bibitem[Illarionov \& Sunyave(1975)]{ill75} Illarionov, A. F. \& Sunyave, R. A., 1975, A\&A, 39, 185
\bibitem[Jiang \& Li(2005)]{jia05} Jiang, Z.-B. \& Li, X.-D. 2005, ChJA\&A, 5, 487
\bibitem[Kaspi, Roberts, \& Harding(2006)]{kas06} Kaspi, V., Roberts, M. S. E., \& Harding, A. K. 2006, in Compact
Stellar X-ray Source, eds. M. van der Klis, W. H. G. Lewin
(Cambridge University Press), in press
\bibitem[Katz(1973)]{kat73} Katz, J. I. 1973, Nat, 246, 87
\bibitem[K\"onigl(1991)]{kon91} K\"onigl, A. 1991, ApJ, 370,
L39
\bibitem[Kramer et al.(2006)]{kra06} Kramer, M., Lyne, A. G. L, O'Brien, J. T., Jordan, C. A., \& Lorimer, D.
R. 2006, Sci, in press
\bibitem[Lai(1999)]{lai99} Lai, D. 1999, ApJ, 524, 1030
\bibitem[Lamb et al.(1985)]{lam85}Lamb, F. K. Shibazaki, N., Alpar, A., \& Shaham, J. 1985, Nat, 317, 681
\bibitem[Lovelace et al.(1995)]{lov95} Lovelace, R. V. E., Romanova, M. M., \& Bisnovatyi-Kogan, G. S. 1995, MNRAS, 275, 244
%\bibitem[Maloney, Begelman, \& Pringle(1996)]{mal96} Maloney, Ph. R., Begelman, M. C. \& Pringle, J. E. 1996, ApJ, 472,
%582
\bibitem[Marsden, Lingenfelter, \& Rothschild(2001)]{mar01} Marsden, D., Lingenfelter, R. E., \& Rothschild, R. E. 2001, ApJ,
547, L45
\bibitem[McLaughlin et al.(2006)]{mcl06} McLaughlin, M. A., Lyne, A. G., Lorimer, D. R., Kramer, M.,
Faulkner, A. J. 2006, Nat, 439, 817
\bibitem[Menou, Perna, \& Hernquist(2001)]{men01} Menou, K., Perna, R., \& Hernquist, L. 2001, ApJ, 554, L63
\bibitem[Michel(1988)]{mic88} Michel, F. C. 1988, Nat, 333, 644
\bibitem[Michel \& Dessler(1981)]{mic81} Michel, F. C. \& Dessler, A. J.
1981, ApJ, 251, 654
\bibitem[Michel \& Dessler(1983)]{mic83} Michel, F. C. \& Dessler, A. J., 1983, Nat, 303, 48
\bibitem[Mineshige et al.(1997)]{min97} Mineshige, S., Nomura, H., Hirose, M., Nomoto, K., \& Suzuki, T.
1997, ApJ, 489, 227
\bibitem[Ogilvie \& Dubus(2001)]{ogi01} Ogilvie, G. I. \& Dubus,
G. 2001, \mnras, 320, 485

\bibitem[Perna \& Hernquist(2000)]{perna00} Perna, R. \&
Hernquist, L. 2000, ApJ, 544, L57
\bibitem[Petterson(1977)]{pet77} Petterson, J. A. 1977, ApJ, 218, 783
\bibitem[Popov et al.(2000)]{pop00} Popov, S. B., Colpi, M., Treves, A., Turolla, R., Lipunov, V. M.,
\& Prokhorov, M.E., 2000, ApJ, 530, 896
\bibitem[Popov et al.(2006)]{pop06} Popov, S. B.,
Turolla, R., \& Possenti, A. 2006, MNRAS, in press
(astro-ph/0603258)
\bibitem[Pringle(1996)]{pri96} Pringle, J. E. 1996, MNRAS, 281, 357
\bibitem[Reich \& F\"urst(1984)]{rei84} Reich, W. \& F\"urst, E. 1984, A\&AS, 57, 165
\bibitem[Reynolds et al.(2006)]{rey06} Reynolds, S. P. et al.
2006, ApJ, 639, L71
\bibitem[Roberts(1974)]{rob74} Roberts, W. J. 1974, ApJ, 187, 575
\bibitem[Ruderman \& Sutherland(1975)]{rud75}Ruderman, M. \& Sutherland, P. G. 1975, ApJ, 196, 51
\bibitem[Shibazaki \& Lamb(1987)]{shi87} Shibazaki, N., \& Lamb, F. K. 1987, ApJ, 318,
767
\bibitem[Sturrock(1971)]{stu71} Sturrock, P. A. 1971, ApJ, 164, 529
\bibitem[Thompson \& Duncan(1995)]{tho95} Thompson, C. \& Duncan,
R. C. 1995, \mnras, 275, 265
\bibitem[Turolla, Possenti, \& Treves(2005)]{tur05} Turolla, R., Possenti, A., \& Treves,
A. 2005, ApJ, 628, L49
\bibitem[Ustyugova et al.(2006)]{ust06} Ustyugova, G. V., Koldoba, A.
V., Romanova, M. M., \& Lovelace, R. V. E. 2006, ApJ, in press
\bibitem[Wang \& Robertson(1985)]{wan85} Wang, Y.-M. \& Robertson, J. A. 1985, A\&A, 151, 361
\bibitem[Wang, Chakrabarty, \& Kaplan(2006)]{wan06} Wang, Z.,
Chakrabarty, D., \& Kaplan, D. L. 2006, Nat, 440, 772
\bibitem[Wang(1983)]{wan83} Wang, Z.-R. 1983, In High energy astrophysics and cosmology
(A85-19326 07-90). Beijing/New York, Science Press/Gordon and
Breach Science Publishers, S.A., p. 270
\bibitem[Yusifov et al.(1995)]{yus95} Yusifov, I. M., Alpar, M. A., Gok, F., \& Huseyinov O. H., 1995,
In The Lives of the Neutron Stars, M. A. Alpar, \"U. Kiziloglu, J.
van Paradijs, eds. (NATO ASI Ser. C, 450; Dordrecht: Kluwer), 201
\bibitem[Zhang \& Gil(2005)]{zha05} Zhang, B. \& Gil, J. 2005, ApJ, 631,
L143
\bibitem[Zhang, Gil \& Dyks(2006)]{zha06} Zhang, B., Gil, J., \& Dyks, J. 2006, ApJ,
submitted (astro-ph/0601063)
\bibitem[Zhu \& Xu(2006)]{zhu06} Zhu, W. W. \& Xu, R. X. 2006, \mnras, 365, L16



\end{thebibliography}
\end{document}